\documentclass[12pt]{article}

\usepackage{a4wide,amsmath,amssymb,bm,graphicx}
\usepackage{jheppub}
\usepackage{slashed}
\usepackage{cancel}
\usepackage{pstricks}
\usepackage{dsfont}

\usepackage[nodisplayskipstretch]{setspace}
\allowdisplaybreaks
\numberwithin{equation}{section}

\newcommand{\ice}[1]{{}}
\newcommand{\EQN}[1]{\hspace{3mm}\fbox{\fbox{$#1$}} \label{#1}}
\newcommand{\mlabel}[1]{ \text{\footnotesize \hspace{3mm}\fbox{\fbox{$#1$}} \label{#1}}}
\renewcommand{\mlabel}[1]{\label{#1}}
\renewcommand{\EQN}[1]{\label{#1}}

\newcommand{\nnb}{\nonumber}
\newcommand{\ed}{\end{document}}
\newcommand{\prd}{\partial}

\newcommand{\beq}{\begin{equation}}
\newcommand{\eeq}{\end{equation}}
\newcommand{\bea}{\begin{eqnarray}}
\newcommand{\eea}{\end{eqnarray}}
\newcommand{\bal}{\[\begin{aligned}}
\newcommand{\eal}{\end{aligned}\]}

\newcommand{\g}{\gamma}
\newcommand{\ga}{\gamma}

\newcommand{\bc}{\begin{center}}
\newcommand{\ec}{\end{center}}
\newcounter{VBQ}

\newcommand{\vv}{v^2}

\newcommand{\F}{\mathbb{G}}

\newcommand{\TT}{{\perp\perp}}
\newcommand{\PT}{{\parallel\perp}}
\newcommand{\NEQN}[1]{\nnb}

\def\bbuildrel#1_#2^#3%
{\mathrel{\mathop{\kern 0pt#1}\limits_{#2}^{#3}}}

\newlength{\blength}
\newlength{\slength}
\newlength{\hlength}
\newlength{\vlength}
\newcommand{\qq}{\bar\psi \psi}
\newcommand{\FF}{FF}

\renewcommand{\qq}{2}
\renewcommand{\FF}{1}

\newcommand{\C}{C}

\title{
\vskip-1cm{\baselineskip14pt
\centerline{\rm\normalsize DESY 21-033\hfill ISSN 0418-9833}
\centerline{\rm\normalsize TTP21-005\hfill}
\centerline{\rm\normalsize March 2021\hfill}}
\vskip1.cm
Operator product expansion of the non-local gluon condensate
}
\author[a]{V. M. Braun,}
\author[b,c]{K. G. Chetyrkin,}
\author[c]{B. A. Kniehl}
\affiliation[a]{
   Institut f\"ur Theoretische Physik, Universit\"at
   Regensburg, 93040~Regensburg, Germany}
\affiliation[b]{Institut f\"ur Theoretische Teilchenphysik, Karlsruhe
  Institute of Technology (KIT), Wolfgang-Gaede-Stra\ss{}e~1, 76131~Karlsruhe, Germany}
\affiliation[c]{
II. Institut f\"ur Theoretische Physik,
Universit\"at  Hamburg, Luruper Chaussee~149, 22761~Hamburg, Germany
}
\emailAdd{vladimir.braun@ur.de}
\emailAdd{Konstantin.Chetyrkin@kit.edu}
\emailAdd{kniehl@desy.de}
\abstract{
We consider the short-distance expansion of the product of two gluon field
strength tensors connected by a straight-line-ordered Wilson line.  The vacuum
expectation value of this nonlocal operator is a common object in studies of
the QCD vacuum structure, whereas its nucleon expectation value is known as the
gluon quasi-parton distribution and is receiving a lot of attention as a tool
to extract gluon distribution functions from lattice calculations.  Extending
our previous study \cite{Braun:2020ymy}, we calculate the three-loop
coefficient functions of the scalar operators in the operator product expansion
up to dimension four. As a by-product, the three-loop anomalous dimension of
the nonlocal two-gluon operator is obtained as well.
}

\begin{document}
\maketitle
\renewcommand{\thefootnote}{\fnsymbol{footnote}}
%%%%%%%%%%%%%%%%%%%%%%%%
\section{Introduction}
%%%%%%%%%%%%%%%%%%%%%%%%

In this work, we construct the operator product expansion (OPE) of the 
non-local two-gluon operator
\begin{align}
 \F_{\mu\nu\alpha\beta}(z) &=  g^2 F_{\mu\nu}(z v) \,[zv,0] \, F_{\alpha\beta}(0)\,, 
\mlabel{F(z)}
\end{align}
where $g$ is the gauge coupling, $F_{\mu\nu}(x)$ is the gluon field strength
tensor, $v^\mu$ is an auxiliary four-vector, with $v^2\ne 0$, and $z$ is a real
number.
In addition, $[zv,0]$ is a straight-line-ordered Wilson line connecting the two field strength tensors,
\begin{align}
 [zv,0] &=  {\cal P} \exp \left[
    ig\, \int_0^z \! d z^\prime\, v^\mu A_{\mu}(z^\prime  v ) \right]\,, 
\mlabel{wline}
\end{align}
with $A_\mu(x)$ being the gluon field in the adjoint representation of the
color gauge group.

The motivation for this study is twofold.
On the one hand, the vacuum expectation value (VEV) of the non-local operator
in Eq.~\eqref{F(z)}, the so-called non-local gluon condensate, describes the correlation of gluon fields in QCD vacuum as a function of their distance and is 
the basic quantity, e.g., in the stochastic model of the QCD vacuum \cite{Dosch:1994wj,DiGiacomo:2000irz}. 
It also governs the effect of gluon condensation on the mass spectra of heavy quarkonia and the short-distance expansion of the 
heavy-quark potential~\cite{Gromes:1982su,Balitsky:1985iw,Campostrini:1986hy,Simonov:1995ui}.
Specifically, it is a central object in nonrelativistic QCD (NRQCD), notably in
potential NRQCD (pNRQCD), where its chromo-electric and chromo-magnetic
components enter the definitions of the heavy-quark potential from the QCD static energy of a heavy quark-antiquark pair and the gluelump masses, the theoretical treatment of quarkonium hybrids \cite{Brambilla:1999xf}, and the determinations of the ultrasoft contribution to the QCD static energy \cite{Brambilla:2006wp} and of long-distrance matrix elements of heavy-quarkonium production \cite{Brambilla:2020ojz} and decay \cite{Brambilla:2020xod}.
Lattice QCD studies of the non-local gluon condensate exist aiming at extracting the gluon correlation length~\cite{DiGiacomo:1992hhp,DElia:1997sdk,Bali:1997aj}, the strong-coupling constant $\alpha_s=g^2/(4\pi)$ via the QCD static energy \cite{Bazavov:2019qoo}, and also its behavior at the deconfinement phase transition at high temperatures (see, e.g., Ref.~\cite{DElia:2002hkf}).     
A similar two-gluon correlator, albeit with a different Wilson line contour, appears in the definition of the rapidity
anomalous dimension (AD), alias Collins--Soper kernel~\cite{Vladimirov:2020umg}. 

On the other hand, the nucleon matrix elements of the same non-local operator, usually referred to as gluon
quasi parton distribution functions (qPDFs), are attracting increasing interest (see Ref.~\cite{Ji:2020ect} for a review).
They can be calculated on the lattice for spacelike separations \cite{Fan:2018dxu} 
and matched to the usual collinear gluon parton distribution functions (PDFs) using continuum perturbation 
theory~\cite{Wang:2017qyg,Wang:2019tgg,Balitsky:2019krf}. This technique is attractive, as it allows one to probe
the gluon PDF more directly than with other approaches, but it is also challenging. In particular, the renormalization 
of the non-local gluon operator involves subtleties \cite{Dorn:1980hs,Dorn:1981wa}, and also lattice calculations
are very challenging due to high statistical noise and the necessity to inject a very large momentum in the 
nucleon, which requires the use of very fine lattices. Using the ratio of the nucleon to vacuum matrix elements 
in such calculations can be advantageous \cite{Braun:2018brg}, as in this way all linear ultraviolet (UV) divergences related to the 
Wilson line renormalization \cite{Dotsenko:1979wb} get canceled.        

In this paper, we consider the OPE of the non-local operator in Eq.~\eqref{F(z)} to three-loop accuracy, taking 
into account all scalar operators up to dimension four. As a by-product of this calculation, 
the three-loop AD (matrix) of the non-local gluon operator is obtained. From the technical 
point of view, this calculation is an extension of our work in Ref.~\cite{Braun:2020ymy}, where the perturbative 
contribution to the OPE was calculated to three-loop accuracy and the two-loop AD was derived. We will mostly adopt
the conventions and the notation of Ref.~\cite{Braun:2020ymy}, a short summary of which is given in Sect.~2. The calculation is described
in Sect.~3. The results for the relevant ADs and coefficient functions (CFs) are presented in Sect.~4.
The renormalization group (RG) evolution equations for the computed CFs as well as the RG  improvement of the purely perturbative contributions are considered  in Sects.~5 and 6.
Section~7 is reserved for a summary and conclusions. 

\section{Preliminaries}

The vacuum expectation value (VEV) of the non-local gluon operator in
Eq.~\eqref{F(z)},
\begin{equation}
\Pi_{\mu\nu\alpha\beta}(z) = 
%\langle 0| g^2 F_{\mu\nu}(zv) \,[zv,0] \, F_{\alpha\beta}(0)|0\rangle\,,
\langle 0|\F_{\mu\nu\alpha\beta}(z)|0\rangle\,,
\end{equation}
can be written in terms of two invariant functions, $\Pi_{\perp\perp}(z)$ and $\Pi_{\parallel\perp}(z)$, which correspond to contributions with different Lorentz symmetry and do not mix under renormalization~\cite{Braun:2020ymy}, as   
\begin{eqnarray}
 \Pi_{\mu\nu\alpha\beta}(z) &=& 
(g^\perp_{\mu\alpha}g^\perp_{\nu\beta} - g^\perp_{\nu\alpha}g^\perp_{\mu\beta}) \Pi_{\perp\perp}(z) 
+ (g^\parallel_{\mu\alpha} g^\perp_{\nu\beta} -  g^\parallel_{\nu\alpha} g^\perp_{\mu\beta}
- g^\parallel_{\mu\beta} g^\perp_{\nu\alpha} +  g^\parallel_{\nu\beta} g^\perp_{\mu\alpha})\Pi_{\parallel\perp}(z)
\nonumber\\
&=&
(g_{\mu\alpha}g_{\nu\beta} - g_{\nu\alpha}g_{\mu\beta})\, \Pi_{\perp\perp}(z)
\nonumber\\
&&{}+  \frac{1}{v^2}( v_\mu v_\alpha g_{\nu\beta} -  v_\nu v_\alpha g_{\mu\beta}
-v_\mu v_\beta g_{\nu\alpha} +  v_\nu v_\beta g_{\mu\alpha})\, [\Pi_{\parallel\perp}(z) - \Pi_{\perp\perp}(z)]\,,   
\mlabel{OPE}
\end{eqnarray}
where
\begin{equation}
   g_{\mu\nu}^\parallel = \frac{v_\mu v_\nu}{v^2}\,, \qquad
   g_{\mu\nu}^\perp = g_{\mu\nu} - \frac{v_\mu v_\nu}{v^2}\,.  
\end{equation}
The renormalization of $\Pi_{\TT}(z)$ and $\Pi_{\PT}(z)$ is determined by their
respective ADs, $\ga_\PT$ and $\ga_\TT$, which are currently known to two-loop accuracy
\cite{Dorn:1980hs,Dorn:1981wa,Braun:2020ymy}. Notice that the renormalization factors are 
local, i.e., they do not depend on the distance between the fields. They can be interpreted as
the renormalization factors of local ``heavy-light'' operators in an effective field theory 
(see Refs.~\cite{Dorn:1980hs,Dorn:1981wa,Braun:2020ymy} for details).

In this work, we consider the OPE in the limit $z\to 0$ of the invariant functions $\Pi_{\TT}(z)$ and $\Pi_{\PT}(z)$
taking into account contributions of the scalar CP-even operators,
\begin{equation}
  O_2=\sum_i m_i \bar{\psi}_i(0)\psi_i(0)\,,
  \qquad
  O_1 = F_{\alpha\beta}(0) F^{\alpha\beta}(0)\,,
  \mlabel{ops}
\end{equation}
where $\psi_i$ stands for the $i$-th quark field with mass $m_i$.
We do not consider operators of mass dimension higher than four.\footnote{%
We also do not consider contributions of tensor operators, e.g., $G_{\alpha\xi}(0)\, G_{\beta}^{\ \xi}(0)$ (symmetrized over the open indices and with the traces subtracted), which do not contribute to the VEV of the non-local operator, but are relevant for hadron matrix elements. Notice that the corresponding OPE in the tree approximation is known through operators of dimension eight \cite{Grozin:1994hd}.
}
To this accuracy, we have
\begin{eqnarray}
  \Pi_{\perp\perp}(z) &    \,\bbuildrel{=\!\!\!=}_{z\to 0\,}^{}  &  \Pi_{\perp\perp}^{m^4}(z)\,\langle\mathds{1}\rangle
  +   C_{\qq}^{\perp\perp}(z) \langle 0| O_{\qq}|0\rangle
+  \frac{g^2}{12} \, C_{\FF}^{\perp\perp}(z)\, \langle 0| O_{\FF}|0\rangle
\,,
%\mlabel{CFtt}
\nonumber\\
\Pi_{\parallel\perp}(z) & \,\bbuildrel{=\!\!\!=}_{z\to 0\,}^{} &  \Pi_{\parallel\perp}^{m^4}(z)\,\langle\mathds{1}\rangle
+  C_{\qq}^{\parallel\perp}(z) \langle 0| O_{\qq}|0\rangle
+  \frac{g^2}{12} \, C_{\FF}^{\parallel\perp}(z)\, \langle 0| O_{\FF}|0\rangle
\,,
\mlabel{CFpt}
\end{eqnarray}
where $\Pi^{m^4}_{\perp\perp}(z)$ and $\Pi^{m^4}_{\parallel\perp}(z)$ stand for the
purely perturbative contributions expanded in the quark masses through order $m_q^4$.
They can be naturally represented as
\begin{eqnarray}
  \Pi^{m^4}_{\TT}(z)&=&\frac{\C^{\TT}_{0}(z)}{z^4}
  +\frac{\C_{m^2}^{\TT}(z)}{z^2}\sum_i m_i^2
  +\C_{m^4,\mathrm{di}}^{\TT}(z)\sum_i m_i^4
  +\C_{m^4,\mathrm{nd}}^{\TT}(z)\sum_{i\not=j} m_i^2\,m_j^2\,,
%\mlabel{Pitt}
\nonumber\\
\Pi^{m^4}_{\PT}(z)&=& \frac{\C^{\PT}_{0}(z)}{z^4}
+\frac{\C_{m^2}^{\PT}(z)}{z^2}\sum_i m_i^2
+\C_{m^4,\mathrm{di}}^{\PT}(z) \sum_i m_i^4
+ \C_{m^4,\mathrm{di}}^{\PT}(z)\sum_{i\not=j} m_i^2\,m_j^2\,,
\mlabel{Pipt}
\end{eqnarray}
where $\Pi_{\TT}^{0}(z)$ and $\Pi_{\PT}^{0}(z)$ correspond to massless,
purely perturbative contributions, 
which are known to two- and three-loop accuracy from Refs.~\cite{Eidemuller:1997bb} and \cite{Braun:2020ymy},
respectively. The new contribution of this work is the calculation of the CFs
$\C_{m^2}^{\TT}$, $\C_{m^4,\mathrm{di}}^{\TT}$, $\C_{m^4,\mathrm{nd}}^{\TT}$, $C_{\qq}^{\TT}$, $C_{\FF}^{\TT}$,
$\C_{m^2}^{\PT}$, $\C_{m^4,\mathrm{di}}^{\PT}$, $\C_{m^4,\mathrm{nd}}^{\PT}$, $C_{\qq}^{\PT}$, $C_{\FF}^{\PT}$,
and the ADs $\ga_\TT$, $\ga_\PT$ to three-loop accuracy.
 
\section{Calculation}

We compute the bare CFs of the operators in Eq.~\eqref{ops} at the three-loop
level using essentially the same techniques as in Ref.~\cite{Braun:2020ymy} 
%with the help of an effective Lagrangian described in \cite{}
and the well-known method of
projectors~\cite{Gorishnii:1983su,Gorishnii:1986gn}. The color factors
are evaluated with the help of the FORM \cite{Vermaseren:2000nd} package
{\tt COLOR} \cite{vanRitbergen:1998pn}.  

Let us briefly discuss the renormalization procedure. 
The renormalization matrix of the operators in Eq.~\eqref{ops} has been known
for a long time \cite{KlubergStern:1974rs,Collins:1976yq,Nielsen:1977sy,Spiridonov:1984br}. Since the CFs $C^\TT_1$ and
$C^\PT_1$ are non-zero already in the tree approximation, their proper
renormalization requires the knowledge of the $Z$ factors, $Z_{\TT}$ and
$Z_{\PT}$, at three loops. Thus, the requirement of finiteness of the
CFs of the operator $ O_1$  provides an alternative way for computing $Z_{\TT}$
and $Z_{\PT}$. From our results for the bare CFs, $(C^\TT_{\FF})_B$ and $(C^\PT_{\FF})_B$, we 
successfully construct the three-loop $Z$ factors, $Z_{\TT}$ and $Z_{\PT}$,
as well as the corresponding ADs. We also find full agreement with 
the corresponding two-loop results, first computed in Ref.~\cite{Braun:2020ymy}.

%The corresponding OPE can be writen as follows:
% \F_{\mu\nu\alpha\beta}(z)

\section{Results}

In this and next two sections, we present our results for the case of standard
QCD with the SU(3) gauge group and $n_f$ active quarks triplets. The results
for the CFs are presented for the case of a spacelike unit vector $v$, with
$v^2=-1$, and the variable $L_z =\ln\left(\mu \mathrm{e}^{\g_E}z/2\right)$ set
to zero. The missing terms proportional to powers $\left(L_z\right)^i$ with
$i=1,2,3$ can be easily restored with the help of the corresponding evolution
equations (see Sect.~\ref{RG}).
\ice{The modifications required for the case of a timelike unit vector
$v$, with $v^2=1$, will be spelled out later.}
Full results for a generic gauge group including the
momentum/position dependence as well the case of $v^2=1$
are appended in the arxiv
submission of this paper as ancillary
files in a computer readable format.

Expanding a generic AD $\ga(a)$ in $a=g^2/(16\pi^2)$ as
\begin{equation}
\ga = \sum_{n \ge 1}
\left(\ga\right)_n a^n
\,,
\end{equation}
our results  for the ADs $\ga_\TT$ and $\ga_\PT$ read:
\begin{eqnarray}
(\ga_{\perp\perp})_1 &=& -3
\,,
\nonumber\\
(\ga_{\perp\perp})_2 &=& -34 +6\, \pi^2 +\frac{13}{3}\, n_f\,,
\nonumber\\
(\ga_{\TT})_3&=&
-\frac{899}{2}-10 \pi^2+18 \pi^4-108  \zeta_{3}
+n_f\, \left(76+\frac{16}{3} \pi^2-2 \pi^4+40  \zeta_{3}\right)
+\frac{2}{3}\, n_f^2
\,,
\nonumber\\
(\ga_{\parallel\perp})_1 &=& 0
\,,
\nonumber\\
(\ga_{\parallel\perp})_2 &=&  6\, \pi^2\,,
\nonumber\\
(\ga_{\PT})_3&=&
8 \pi^2+18 \pi^4
+n_f\, \left(\frac{22}{3} \pi^2-2 \pi^4\right)
\,.
\EQN{gpt3}
\end{eqnarray} 

Expanding the CFs $\C_{m^2}$, $\C_{m^4,\mathrm{di}}$, $\C_{m^4,\mathrm{nd}}$, $C_{\qq}$ and $C_{\FF}$ as
\begin{gather}
  \C_{m^2}          = \sum_{n \ge 1} \left(\C_{m^2}\right)_n a^n\,, \qquad 
  \C_{m^4,\mathrm{di}} = \sum_{n \ge 2} \left(\C_{m^4,\mathrm{di}}\right)_n a^n\,, \qquad 
  \C_{m^4,\mathrm{nd}} = \sum_{n \ge 3} \left(\C_{m^4,\mathrm{nd}} \right)_n a^n\,,
\nonumber\\
  C_2 =  \sum_{n \ge 1} g^2\, \left(C_2\right)_n a^n\,, \qquad 
  C_1 =  1 +\sum_{n \ge 1} \left(C_1\right)_n a^n\,,
\EQN{CFs}  
\end{gather}
we find the coefficients appearing in Eq.~\eqref{CFs} to be
\begin{eqnarray}
 \left(\C_{m^2}^{\TT}\right)_2&=&
 128
\,,
\nonumber\\
\left(\C_{m^2}^{\TT}\right)_3&=&
\frac{15872}{3}+\frac{256}{3} \pi^2+\frac{128}{45} \pi^4+768  \zeta_{3}
-\frac{1024}{9}\, n_f
\,,
\nonumber\\
\left(\C_{m^2}^{\PT}\right)_2&=&0
\,,
\nonumber\\
\left(\C_{m^2}^{\PT}\right)_3&=&
3840+768 \pi^2-\frac{128}{3} \pi^4
-\frac{512}{3}\, n_f
\,,
\nonumber\\
\left(\C_{m^4,\mathrm{di}}^{\TT}\right)_2&=&
-\frac{28}{3}
\,,
\nonumber\\
\left(\C_{m^4,\mathrm{di}}^{\TT}\right)_3&=&
\frac{3650}{27}-\frac{176}{9} \pi^2-\frac{16}{135} \pi^4-\frac{4384}{9}  \zeta_{3}
+n_f\, \left(-\frac{412}{9}+\frac{128}{3}  \zeta_{3}\right)
\,,
\nonumber\\
\left(\C_{m^4,\mathrm{di}}^{\PT}\right)_2&=&
\frac{20}{3}
\,,
\nonumber\\
\left(\C_{m^4,\mathrm{di}}^{\PT}\right)_3&=&
-\frac{11566}{27}+\frac{800}{9} \pi^2-\frac{16}{135} \pi^4-\frac{6880}{9}  \zeta_{3}
+n_f\, \left(-\frac{68}{3}+\frac{128}{3}  \zeta_{3}\right)
\,,
\nonumber\\
\left(\C_{m^4,\mathrm{nd}}^{\TT}\right)_2&=&0
\,,
\nonumber\\
\left(\C_{m^4,\mathrm{nd}}^{\TT}\right)_3&=&
\frac{32}{9}
\,,
\nonumber\\
\left(\C_{m^4,\mathrm{nd}}^{\PT}\right)_2&=&0
\,,
\nonumber\\
 \left(\C_{m^4,\mathrm{nd}}^{\PT}\right)_3&=&
-\frac{544}{9}
\,,
\nonumber\\
\left(C_2^{\TT}\right)_1&=&
-\frac{8}{9}
\,,
\nonumber\\
\left(C_2^{\TT}\right)_2&=&
-\frac{175}{27}-\frac{52}{9} \pi^2
-2\, n_f
\,,
\nonumber\\
\left(C_2^{\TT}\right)_3&=&
-\frac{36190}{81}+\frac{56222}{243} \pi^2-\frac{346}{15} \pi^4+\frac{8996}{27}  \zeta_{3}-\frac{848}{3} \pi^2  \zeta_{3}+1400  \zeta_{5}
\nonumber\\
&&{}+n_f\, \left(-\frac{7654}{243}+\frac{8}{5} \pi^4-\frac{928}{9}  \zeta_{3}\right)
+n_f^2\, \left(-\frac{1198}{729}+\frac{32}{9}  \zeta_{3}\right)
\,,
\nonumber\\
\left(C_2^{\PT}\right)_1&=&
\frac{4}{9}
\,,
\nonumber\\
\left(C_2^{\PT}\right)_2&=&
\frac{725}{27}-\frac{28}{3} \pi^2
-\frac{62}{27}\, n_f
\,,
\nonumber\\
\left(C_2^{\PT}\right)_3&=&
\frac{176317}{81}-\frac{235214}{243} \pi^2-\frac{3254}{81} \pi^4+4620  \zeta_{3}+\frac{1904}{3} \pi^2  \zeta_{3}-3320  \zeta_{5}
\nonumber\\
&&{}+n_f\, \left(-\frac{54658}{243}+24 \pi^2+\frac{8}{5} \pi^4-\frac{2176}{9}  \zeta_{3}\right)
+n_f^2\, \left(\frac{458}{729}+\frac{32}{9}  \zeta_{3}\right)
\,,
\nonumber\\
\left(C_1^{\TT}\right)_1&=&
\frac{27}{2}
\,,
\nonumber\\
\left(C_1^{\TT}\right)_2&=&
\frac{5737}{12}+54 \pi^2-144  \zeta_{3}
-\frac{160}{9}\, n_f
\,,
\nonumber\\
\left(C_1^{\TT}\right)_3&=&
\frac{755000}{27}+\frac{4295}{2} \pi^2+\frac{3436}{15} \pi^4-16176  \zeta_{3}-1176 \pi^2  \zeta_{3}+1800  \zeta_{5}
\nonumber\\
&&{}+n_f\, \left(-\frac{321653}{108}-\frac{866}{9} \pi^2-\frac{848}{45} \pi^4+\frac{20446}{27}  \zeta_{3}-16 \pi^2  \zeta_{3}+600  \zeta_{5}\right)
\nonumber\\
&&{}+n_f^2\, \left(\frac{5542}{81}-\frac{16}{3}  \zeta_{3}\right)
\,,
\nonumber\\
\left(C_1^{\PT}\right)_1&=&
\frac{21}{2}
\,,
\nonumber\\
\left(C_1^{\PT}\right)_2&=&
\frac{1321}{4}+61 \pi^2-144  \zeta_{3}
-\frac{41}{3}\, n_f
\,,
\nonumber\\
\left(C_1^{\PT}\right)_3&=&
\frac{69761}{3}+\frac{9125}{3} \pi^2+\frac{1242}{5} \pi^4-18168  \zeta_{3}-1968 \pi^2  \zeta_{3}+6120  \zeta_{5}
\nonumber\\
&&{}+n_f\, \left(-\frac{820793}{324}-\frac{1240}{9} \pi^2-\frac{928}{45} \pi^4+\frac{21082}{27}  \zeta_{3}-16 \pi^2  \zeta_{3}+600  \zeta_{5}\right)
\nonumber\\
&&{}+\frac{15025}{243}\, n_f^2
\,.
\mlabel{xtCO1ptL3}
\end{eqnarray} 
Numerically, for $n_f=3$, we obtain
\begin{eqnarray}
\left(\C_{m^2}^{\TT}\right)_{n_f=3}&=&
8 \, a^2 \left[ 1 
+ \,a \,\left( 13.656 + 10.000  L_z  \right)\right]
\,,
\nonumber\\
\left(\C_{m^2}^{\PT}\right)_{n_f=3}&=&
105.496 \, \,a^3
\,,
\nonumber\\
\left(\C_{m^4,\mathrm{di}}^{\TT}\right)_{n_f=3}&=&
-\frac{7}{12} \, a^2 \left[ 1 - 3.429  L_z  + 3.429  \, L_z^2\right.
\nonumber\\
&&{}+\left. \,a \,\left( 17.099 - 41.523  L_z  + 13.000  \, L_z^2+ 19.429  \, L_z^3\right)\right]
\,,
\nonumber\\
\left(\C_{m^4,\mathrm{di}}^{\PT}\right)_{n_f=3}&=&
\frac{5}{12} \, a^2 \left[ 1 - 4.800  \, L_z^2
+ \,a \,\left(- 14.837 + 89.515  L_z  - 62.600  \, L_z^2- 41.600  \, L_z^3\right)\right]
\,,
\nonumber\\
\left(\C_{m^4,\mathrm{nd}}^{\TT}\right)_{n_f=3}&=&
 \,a^3\, \left( 0.056 - 2.000  L_z  \right)
\,,
\nonumber\\
\left(\C_{m^4,\mathrm{nd}}^{\PT}\right)_{n_f=3}&=&
 \,a^3\, \left(- 0.944 - 2.000  L_z  \right)
\,,
\nonumber\\
\left(C_2^{\TT}\right)_{n_f=3}&=&
-\frac{2}{9} \,a \left[ 1 - 3.000  L_z  
+ \,a \,\left( 19.549 - 26.875  L_z  - 4.500  \, L_z^2\right)\right.
\nonumber\\
&&{}+ \left.\,a^2\, \left( 132.651 - 104.864  L_z  - 155.226  \, L_z^2- 13.500  \, L_z^3\right)\right]
\,,
\nonumber\\
\left(C_2^{\PT}\right)_{n_f=3}&=&
\frac{1}{9} \,a \left[ 1 + 6.000  L_z  
+ \,a \,\left(- 40.586 + 70.250  L_z  + 27.000  \, L_z^2\right)\right.
\nonumber\\
&&{}+\left. \,a^2\, \left(- 277.555 + 95.216  L_z  + 622.451  \, L_z^2+ 121.500  \, L_z^3\right)\right]
\,,
\nonumber\\
\left(C_1^{\TT}\right)_{n_f=3}&=&
 1 
+ \,a \,\left( 3.375 - 3.000  L_z  \right)
+ \,a^2\, \left( 49.038 + 6.617  L_z  - 2.250  \, L_z^2\right)
\nonumber\\
&&{}+ \,a^3\, \left( 425.570 + 350.427  L_z  + 5.519  \, L_z^2- 4.500  \, L_z^3\right)
\,,
\nonumber\\
\left(C_1^{\PT}\right)_{n_f=3}&=&
 1 
+ 2.625 \, \,a 
+ \,a^2\, \left( 44.887 + 18.617  L_z  \right)
\nonumber\\
&&{}+ \,a^3\, \left( 399.885 + 494.172  L_z  + 83.776  \, L_z^2\right)
\,.
\EQN{xtCO1ptL_nf=3}
 \end{eqnarray} 

\boldmath
\section{RG improvements of $\Pi_\TT^0$ and $\Pi_\PT^0$}
\unboldmath

In general, a multiplicatively renormalizable structure function $\Pi$
depends on both a renormalization prescription, or scheme, and a normalization scale $\mu$. It is 
convenient to deal with the scheme and scale invariant version of $\Pi$,
which we denote as $\hat{\Pi}$. Given the RG equation for $\Pi$, 
 \begin{equation} 
\mu^2 \frac{\mathrm{d}}{\mathrm{d} \mu^2}
\Pi(a,\mu) \equiv
\left(
\mu^2\frac{\partial}{\partial\mu^2}
 +
\beta(a)
a
\frac{\partial}{\partial a}
\right)
      \Pi(a,\mu)          = \gamma(a)\, \Pi(a,\mu) 
\mlabel{rg:Pi2}
\,,
 \end{equation} 
a formal solution for $\hat{\Pi}$ reads
 \begin{equation}  
\hat{\Pi}(a)  = 
 \frac{\Pi(a,\mu)}{f(a)}\,,\qquad
f(a)  = 
\exp\int^a \,\frac{ \mathrm{d} x}{x}\ \frac{\gamma(x)}{\beta(x)}\,.
\mlabel{formal_sol}
 \end{equation}
Through the order of interest here, we have
\begin{eqnarray}  
f(a) &=& (a)^{\bar{ \gamma}_1}  \left\{ 1 + (\bar{ \gamma_2} - \bar{\beta_2}\bar{ \gamma _1})a
\right.
\nonumber\\
&&{}+
\frac{1}{2}
\left.
\left[
(\bar{\gamma} _2 - \bar{\beta}_2\bar{\gamma}_1)^2
+
\bar{\gamma}_3 + \bar{\beta}_2^2\bar{\gamma}_1
- \bar{\beta}_2\bar{\gamma}_2 -\bar{\beta}_3\bar{\gamma}_1
\right] a^2
 + {\cal O}(a^3)
\right\}
{},
\mlabel{c(a)}
\end{eqnarray} 
where $\bar{\gamma}_i =  \gamma _i/\beta_1$, $\bar{\beta}_i = \beta_i/\beta_1$
$(i=1,2,3)$, and the coefficients $\beta_i$ of the beta function are defined as
 \begin{equation} 
\beta(a) =
\mu^2  \frac{ \mathrm d}{ \mathrm d \mu^2} \ln a
=
 \sum_{n=1}^\infty \beta_{n} a^{n}
\,,
\mlabel{beta:function:generic}
 \end{equation} 
with $\beta_1 = -11 + \frac{2}{3}\,n_f$, etc.

 \newcommand{\hPi}{\hat{\Pi}}
 
 \newcommand{\hF}{\hat{F}}

 Our results for the scheme invariant functions $\hPi^0_\TT$ and $\hPi^0_\PT$ in
 position space are given by 
\begin{eqnarray}
\hPi^0_\TT(z>0, \vv=-1) &=&  \frac{128}{ z^4}
%a^{1-\frac{\left(\ga_\TT\right)_1}{\beta_1}}
a^{1 + 6/\beta_1}
\left[ 1 +  \sum_{n=1}^2 \left( \hF_\TT \right)_n  a^n\right]
\,,
%\mlabel{Ftt:z}
\nonumber\\
\hPi^0_\PT(z>0, \vv=-1) &=&  -\frac{128 \, a}{ z^4}
\left[ 1 +  \sum_{n=1}^2 \left( \hF_\PT \right)_n  a^n\right]
\,.
\mlabel{Fpt:z}
\end{eqnarray}
In phenomenological applications, one usually has $n_f =0,2,3$.
The CFs in Eq.~\eqref{Fpt:z} then take the values
\begin{eqnarray}
 \left(\hF_\TT\right)_1^{n_f =0}&=&
\frac{9151}{363}+\frac{56}{11} \pi^2
+10\, L_z  
\,,
%\EQN{hzTT.a1nf0}
\nonumber\\
\left(\hF_\TT\right)_2^{n_f =0}&=&
\frac{265666457}{263538}+\frac{808070}{3993} \pi^2+\frac{798}{121} \pi^4-\frac{10206}{11}  \zeta_{3}
\nonumber\\
&&{}+ L_z  \,\left(\frac{326492}{363}+\frac{1792}{11} \pi^2\right)
+160\, L_z  ^2
\,,
%\EQN{hzTT.a2nf0}
\nonumber\\
\left(\hF_\TT\right)_1^{n_f =2}&=&
\frac{190177}{7569}+\frac{152}{29} \pi^2
+\frac{22}{3}\, L_z  
\,,
%\EQN{hzTT.a1nf2}
\nonumber\\
 \left(\hF_\TT\right)_2^{n_f =2}&=&
\frac{102303720467}{114579522}+\frac{45401582}{219501} \pi^2+\frac{6042}{841} \pi^4-\frac{71998}{87}  \zeta_{3}
\nonumber\\
&&{}+ L_z  \,\left(\frac{16534820}{22707}+\frac{12160}{87} \pi^2\right)
+\frac{880}{9}\, L_z  ^2
\,,
%\EQN{hzTT.a2nf2}
\nonumber\\
\left(\hF_\TT\right)_1^{n_f =3}&=&
\frac{677}{27}+\frac{16}{3} \pi^2
+6\, L_z  
\,,
%\EQN{hzTT.a1nf3}
\nonumber\\
\left(\hF_\TT\right)_2^{n_f =3}&=&
\frac{1216447}{1458}+\frac{16982}{81} \pi^2+\frac{68}{9} \pi^4-\frac{2330}{3}  \zeta_{3}
\nonumber\\
&&{}+ L_z  \,\left(\frac{5800}{9}+128 \pi^2\right)
+72\, L_z  ^2
\,,
%\EQN{hzTT.a2nf3}
\nonumber\\
\left(\hF_\PT\right)_1^{n_f =0}&=&
\frac{28}{3}+\frac{56}{11} \pi^2
+22\, L_z  
\,,
%\EQN{hzPT.a1nf0}
\nonumber\\
\left(\hF_\PT\right)_2^{n_f =0}&=&
\frac{9011}{18}+\frac{23299}{363} \pi^2+\frac{798}{121} \pi^4-1026  \zeta_{3}
\nonumber\\
&&{}+ L_z  \,\left(\frac{1844}{3}+224 \pi^2\right)
+484\, L_z  ^2
\,,
%\EQN{hzPT.a2nf0}
\nonumber\\
\left(\hF_\PT\right)_1^{n_f =2}&=&
\frac{88}{9}+\frac{152}{29} \pi^2
+\frac{58}{3}\, L_z  
\,,
%\EQN{hzPT.a1nf2}
\nonumber\\
\left(\hF_\PT\right)_2^{n_f =2}&=&
\frac{74591}{162}+\frac{567709}{7569} \pi^2+\frac{6042}{841} \pi^4-\frac{2798}{3}  \zeta_{3}
\nonumber\\
&&{}+ L_z  \,\left(\frac{14348}{27}+\frac{608}{3} \pi^2\right)
+\frac{3364}{9}\, L_z  ^2
\,,
%\EQN{hzPT.a2nf2}
\nonumber\\
\left(\hF_\PT\right)_1^{n_f =3}&=&
10+\frac{16}{3} \pi^2
+18\, L_z  
\,,
%\EQN{hzPT.a1nf3}
\nonumber\\
\left(\hF_\PT\right)_2^{n_f =3}&=&
\frac{2627}{6}+\frac{2185}{27} \pi^2+\frac{68}{9} \pi^4-886  \zeta_{3}
\nonumber\\
&&{}+ L_z  \,\left(488+192 \pi^2\right)
+324\, L_z  ^2
\,.
\EQN{hzPT.a2nf3}
\end{eqnarray} 
Numerically, the same results read
\begin{eqnarray}
\left(\hF_{\TT}\right)_{n_f=0}&=&
 1 
+ \,a \,\left( 18.864 + 2.500  L_z  \right)
+ \,a^2\, \left( 158.283 + 156.705  L_z  + 10.000  \, L_z^2\right)
\,,
%\EQN{zhFttL_nf=0}
\nonumber\\
\left(\hF_{\TT}\right)_{n_f=2}&=&
 1 
+ \,a \,\left( 19.214 + 1.833  L_z  \right)
+ \,a^2\, \left( 164.958 + 131.729  L_z  + 6.111  \, L_z^2\right)
\,,
%\EQN{zhFttL_nf=2}
\nonumber\\
\left(\hF_{\TT}\right)_{n_f=3}&=&
 1 
+ \,a \,\left( 19.428 + 1.500  L_z  \right)
+ \,a^2\, \left( 169.120 + 119.235  L_z  + 4.500  \, L_z^2\right)
\,,
%\EQN{zhFttL_nf=3}
\nonumber\\
\left(\hF_{\PT}\right)_{n_f=0}&=&
 1 
+ \,a \,\left( 14.895 + 5.500  L_z  \right)
+ \,a^2\, \left( 33.950 + 176.591  L_z  + 30.250  \, L_z^2\right)
\,,
%\EQN{zhFptL_nf=0}
\nonumber\\
\left(\hF_{\PT}\right)_{n_f=2}&=&
 1 
+ \,a \,\left( 15.377 + 4.833  L_z  \right)
+ \,a^2\, \left( 48.713 + 158.228  L_z  + 23.361  \, L_z^2\right)
\,,
%\EQN{zhFptL_nf=2}
\nonumber\\
\left(\hF_{\PT}\right)_{n_f=3}&=&
 1 
+ \,a \,\left( 15.659 + 4.500  L_z  \right)
+ \,a^2\, \left( 56.719 + 148.935  L_z  + 20.250  \, L_z^2\right)
\,.\qquad
\EQN{zhFptL_nf=3}
\end{eqnarray} 
 
\section{RG improvements of the CFs \label{RG}}

The ${\cal O}(1/z^4)$ and ${\cal O}(1/z^2)$ terms in Eq.~\eqref{Pipt},
\begin{equation}
\Pi^{m^2}_{\TT}(z) =  \frac{C_{0}^{\TT}(z)}{z^4}  +  \frac{C_{m^2}^{\TT}(z)}{z}\sum_i m_i^2
 \,,
\end{equation}  
satisfy a simple evolution equation of the form\footnote{%
  In this section, we only consider CFs corresponding to the invariant function $\Pi_{\TT}$. The corresponding relations for $\Pi_{\PT}$ emerge by replacing $\TT$ with $\PT$.}
\begin{equation}
  \mu^2\frac{d }{d \mu^2}\, \Pi^{m^2}_{\TT}(z) = \ga_{\TT}(a) \,  \Pi^{m^2}_{\TT}(z),
\EQN{RGm2}
\end{equation}
where
\begin{equation}
\mu^2\frac{d }{d \mu^2} \equiv 
\mu^2\frac{\prd }{\prd \mu^2} + \beta(a)\, a\, \frac{\prd }{\prd a}
+
\ga_m(a) \,  m_i \, \frac{\prd }{\prd m_i}
\,,
\end{equation}
with $\ga_m(a) = - 4\, a  + (-\frac{202}{3} + \frac{20}{9}\,n_f)\,a^2  + \cdots$ being the quark mass anomalous dimension.

The evolution equations for the remaining CFs are  more complicated 
due to mixing between the operators $O_1$, $O_2$ and combinations quartic in
quark masses \cite{KlubergStern:1974rs,Spiridonov:1984br,Spiridonov:1988md}.
The mixing is described by 
\begin{eqnarray}
\mu^2\frac{d }{d \mu^2} O_1 &=&
\ga_{11}\, O_1
+ \ga_{12}\, O_2
+ 4a\frac{ \prd}{\prd a}\g_0\,,
\nonumber\\
\mu^2\frac{d }{d \mu^2} O_2 &=& -4\, \ga_0
%- m_{i}
%\frac{\prd }{\prd m_i} \g_0
\,,
\EQN{anom.dim4}
\end{eqnarray}
where
\begin{equation}
  \ga_{11} = - a\frac{ \prd}{\prd a}\beta\,, \qquad
  \ga_{12} =  4 a\, \frac{ \prd}{\prd a}\g_m\,, 
\end{equation}
and $\ga_0$ is the anomalous dimension of the vacuum energy
\cite{Spiridonov:1988md,Chetyrkin:1994ex,Chetyrkin:2018avf},
\begin{equation}
\g_0(a,m) = \g_0^{\mathrm{di}}(a) \sum_i m_i^4
+
\g_0^{\mathrm{nd}}(a) \sum_{ i\not= j} m_i^2 m_j^2
\,,
\end{equation}
with
\begin{eqnarray}
\g_0^{\mathrm{di}}(a) &=&
-\frac{3}{16\pi^2}
\left[  1 + \frac{16}{3} a
 + \left(\frac{626}{9}- \frac{32}{3}\zeta(3) - \frac{20}{3} n_f \right)a^2
\right]
\,,
\nonumber\\
\g_0^{\mathrm{nd}}(a) &=& \frac{6}{\pi^2}a^2\,.
\EQN{vac.energy}
\end{eqnarray}
The resulting evolution equations for the CFs
$\mathds{C_1}\equiv \frac{g^2}{12} C_1$,
$C_2$, $C_{m^4,\mathrm{di}}$, and $C_{m^4,\mathrm{nd}}$ read:
\begin{eqnarray}
\mu^2\frac{d }{d \mu^2} \,\mathds{C}_1^{\TT} &=& (\ga_{\TT} - \ga_{11})\, \mathds{C}_1^{\TT}\,,
\nonumber\\
\mu^2\frac{d }{d \mu^2} \,C_2^{\TT} &=& \ga_{\TT} \, C_2^{\TT} - \ga_{12}\,\mathds{C}_1^{\TT} 
\,,
\nonumber\\
\mu^2\frac{d }{d \mu^2}\, C_{m^4,\mathrm{di}}^{\TT} &=& (\ga_{\TT}  -4\, \ga_m)\,
C_{m^4,\mathrm{di}}^{\TT}
- \left(4a\,\frac{ \prd}{\prd a}\,\g_0^{\mathrm{di}}\right)\, \mathds{C}_1^{\TT} 
+ 4\, C_2^{\TT}\, \ga_0^{\mathrm{di}}
\,,
\nonumber\\
\mu^2\frac{d }{d \mu^2}\, C_{m^4,\mathrm{nd}}^{\TT} &=& (\ga_{\TT}  -4\, \ga_m)\,
C_{m^4,\mathrm{nd}}^{\TT}
- \left(4a\,\frac{ \prd}{\prd a}\,\g_0^{\mathrm{nd}}\right)\, \mathds{C}_1^{\TT} 
+ 4\, C_2^{\TT}\, \ga_0^{\mathrm{nd}}
\,.
\EQN{RGrest}
\end{eqnarray}
\noindent
We checked\footnote{
  In addition to the ADs $\ga_\TT, \ga_\PT$, and $ \ga_0$, we have used the three-loop $\beta$ function \cite{Tarasov:1980au,Larin:1993tp} and the two-loop quark mass anomalous dimension $\g_m$ \cite{Tarrach:1980up,Nachtmann:1981zg}.}
  that our results do satisfy Eqs.~\eqref{RGm2} and \eqref{RGrest}.
 
\section{Conclusions}

We considered the non-local operator consisting of two gluon field strength
tensors connected by a straight Wilson line and studied its VEV, which is
determined by two nonperturbative Lorentz scalar functions.
For each of the latter, we performed an OPE through mass dimension four, which,
besides a purely perturbative structure function, involves the VEVs of three
local operators, and calculated the AD of the structure function and the CFs of
the three local operators through three loops, order $a^3$, in the modified
minimal-subtraction ($\overline{\mathrm{MS}}$) scheme.
In our recent work \cite{Braun:2020ymy}, the structure functions were derived
through three loops and the ADs through two loops.
Using the ADs, we also performed a RG improvement of the structure functions
and presented their renormalization scheme invariant counterparts thus
resulting, again through three loops.
%These results can be immediately used by the lattice QCD community to advance
%studies of the QCD vacuum structure and render extractions of gluon qPDFs of
%protons and nuclei more reliable.
These results should be of interest to the lattice QCD community in studies of
the QCD vacuum structure and serve as normalization factors in
calculations of gluon PDFs within the qPDF approach, eliminating the need for
nonperturbative subtractions of linear divergences.

\section*{Acknowledgments}

The work of V.M.B. and B.A.K. was supported in part by DFG 
Research Unit FOR 2926 under Grant No.\ 409651613.
The work of K.G.C. was supported by DFG Grant No.\ CH~1479/2-1. 
%

%\bibliography{opeGG}

\begin{thebibliography}{10}

\bibitem{Braun:2020ymy}
V.~M.~Braun, K.~G.~Chetyrkin and B.~A.~Kniehl, \emph{{Renormalization of parton
  quasi-distributions beyond the leading order: spacelike vs.\ timelike}},
  \href{http://dx.doi.org/10.1007/JHEP07(2020)161}{\emph{JHEP} {\bfseries 07}
  (2020) 161}, [\href{https://arxiv.org/abs/2004.01043}{{\ttfamily
  2004.01043 [hep-ph]}}].

\bibitem{Dosch:1994wj}
H.~G. Dosch, \emph{{Nonperturbative methods in quantum chromodynamics}},
  \href{http://dx.doi.org/10.1016/0146-6410(94)90044-2}{\emph{Prog. Part. Nucl.
  Phys.} {\bfseries 33} (1994) 121--199}.

\bibitem{DiGiacomo:2000irz}
A.~Di~Giacomo, H.~G. Dosch, V.~I. Shevchenko and {\relax Yu}.~A. Simonov,
  \emph{{Field correlators in QCD. Theory and applications}},
  \href{http://dx.doi.org/10.1016/S0370-1573(02)00140-0}{\emph{Phys. Rept.}
  {\bfseries 372} (2002) 319--368},
  [\href{https://arxiv.org/abs/hep-ph/0007223}{{\ttfamily hep-ph/0007223}}].

\bibitem{Gromes:1982su}
D.~Gromes, \emph{{Space-time dependence of the gluon condensate correlation
  function and quarkonium spectra}},
  \href{http://dx.doi.org/10.1016/0370-2693(82)90397-5}{\emph{Phys. Lett. B}
  {\bfseries 115} (1982) 482--486}.

\bibitem{Balitsky:1985iw}
I.~I.~Balitsky, \emph{{Wilson loop for the stretched contours in vacuum fields and
  the small-distance behaviour of the interquark potential}},
  \href{http://dx.doi.org/10.1016/0550-3213(85)90215-9}{\emph{Nucl. Phys. B}
  {\bfseries 254} (1985) 166--186}.

\bibitem{Campostrini:1986hy}
M.~Campostrini, A.~Di~Giacomo and {\^S}.~Olejnik, \emph{{On the possibility of
  detecting gluon condensation from the spectra of heavy quarkonia}},
  \href{http://dx.doi.org/10.1007/BF01551081}{\emph{Z. Phys. C} {\bfseries 31}
  (1986) 577--582}.

\bibitem{Simonov:1995ui}
{\relax Yu}.~A.~Simonov, S.~Titard and F.~J.~Yndur\'ain, \emph{{Heavy quarkonium systems and
  nonperturbative field correlators}},
  \href{http://dx.doi.org/10.1016/0370-2693(95)00609-O}{\emph{Phys. Lett. B}
  {\bfseries 354} (1995) 435--441},
  [\href{https://arxiv.org/abs/hep-ph/9504273}{{\ttfamily hep-ph/9504273}}].

\bibitem{Brambilla:1999xf}
  N.~Brambilla, A.~Pineda, J.~Soto and A.~Vairo, \emph{{Potential NRQCD: an
  effective theory for heavy quarkonium}},
  \href{https://doi.org/10.1016/S0550-3213(99)00693-8}{\emph{Nucl. Phys. B}
  {\bfseries 566} (2000) 275--310},
  [\href{https://arxiv.org/abs/hep-ph/9907240}{{\ttfamily hep-ph/9907240}}].

\bibitem{Brambilla:2006wp}
  N.~Brambilla, X.~Garcia i Tormo, J.~Soto and A.~Vairo, \emph{{The logarithmic
  contribution to the QCD static energy at N$^4$LO}},
  \href{https://doi.org/10.1016/j.physletb.2007.02.015}{\emph{Phys. Lett. B}
  {\bfseries 647} (2007) 185--193},
  [\href{https://arxiv.org/abs/hep-ph/0610143}{{\ttfamily hep-ph/0610143}}].

\bibitem{Brambilla:2020ojz}
  N.~Brambilla, H.~S.~Chung and A.~Vairo, \emph{{Inclusive Hadroproduction of
  $P$-Wave Heavy Quarkonia in Potential Nonrelativistic QCD}},
  \href{https://doi.org/10.1103/PhysRevLett.126.082003}{\emph{Phys. Rev. Lett.}
  {\bfseries 126} (2021) 082003},
  [\href{https://arxiv.org/abs/2007.07613}{{\ttfamily 2007.07613 [hep-ph]}}].

\bibitem{Brambilla:2020xod}
  N.~Brambilla, H.~S.~Chung, D.~M\"uller and A.~Vairo, \emph{{Decay and
  electromagnetic production of strongly coupled quarkonia in pNRQCD}},
  \href{https://doi.org/10.1007/JHEP04(2020)095}{\emph{JHEP} {\bfseries 04}
  (2020) 095},
  [\href{https://arxiv.org/abs/2002.07462}{{\ttfamily 2002.07462 [hep-ph]}}].

\bibitem{DiGiacomo:1992hhp}
A.~Di~Giacomo and H.~Panagopoulos, \emph{{Field strength correlations in the
  QCD vacuum}},
  \href{http://dx.doi.org/10.1016/0370-2693(92)91311-V}{\emph{Phys. Lett. B}
  {\bfseries 285} (1992) 133--136}.

\bibitem{DElia:1997sdk} %30.5.1997
M.~D'Elia, A.~Di~Giacomo and E.~Meggiolaro, \emph{{Field strength correlators
  in full QCD}},
  \href{http://dx.doi.org/10.1016/S0370-2693(97)00814-9}{\emph{Phys. Lett. B}
  {\bfseries 408} (1997) 315--319},
  [\href{https://arxiv.org/abs/hep-lat/9705032}{{\ttfamily hep-lat/9705032}}].

\bibitem{Bali:1997aj} %25.9.1997
  G.~S.~Bali, N.~Brambilla and A.~Vairo, \emph{{A lattice determination of QCD
  field strength correlators}},
  \href{https://doi.org/10.1016/S0370-2693(97)01583-9}{\emph{Phys.\ Lett.\ B}
  {\bfseries 421} (1998) 265--272},
  [\href{https://arxiv.org/abs/hep-lat/9709079}{{\ttfamily hep-lat/9709079}}].

\bibitem{Bazavov:2019qoo}
  A.~Bazavov, N.~Brambilla, X.~Garcia i Tormo, P.~Petreczky, J.~Soto, A.~Vairo
  and J.~H.~Weber (TUMQCD Collaboration), \emph{{Determination of the QCD
  coupling from the static energy and the free energy}},
  \href{https://doi.org/10.1103/PhysRevD.100.114511}{\emph{Phys. Rev. D}
  {\bfseries 100} (2019) 114511},
  [\href{https://arxiv.org/abs/1907.11747}{{\ttfamily 1907.11747 [hep-lat]}}].

\bibitem{DElia:2002hkf}
M.~D'Elia, A.~Di~Giacomo and E.~Meggiolaro, \emph{{Gauge invariant field
  strength correlators in pure Yang-Mills and full QCD at finite temperature}},
  \href{http://dx.doi.org/10.1103/PhysRevD.67.114504}{\emph{Phys. Rev. D}
  {\bfseries 67} (2003) 114504},
  [\href{https://arxiv.org/abs/hep-lat/0205018}{{\ttfamily hep-lat/0205018}}].

\bibitem{Vladimirov:2020umg}
A.~A.~Vladimirov, \emph{{Self-Contained Definition of the Collins-Soper
  Kernel}}, \href{http://dx.doi.org/10.1103/PhysRevLett.125.192002}{\emph{Phys.
  Rev. Lett.} {\bfseries 125} (2020) 192002},
  [\href{https://arxiv.org/abs/2003.02288}{{\ttfamily 2003.02288 [hep-ph]}}].

\bibitem{Ji:2020ect}
X.~Ji, Y.-S. Liu, Y.~Liu, J.-H. Zhang and Y.~Zhao, \emph{{Large-Momentum
  Effective Theory}},  \href{https://arxiv.org/abs/2004.03543}{{\ttfamily
  2004.03543 [hep-ph]}}.

\bibitem{Fan:2018dxu}
Z.-Y. Fan, Y.-B. Yang, A.~Anthony, H.-W. Lin and K.-F. Liu, \emph{{Gluon
  Quasi-Parton-Distribution Functions from Lattice QCD}},
  \href{http://dx.doi.org/10.1103/PhysRevLett.121.242001}{\emph{Phys. Rev.
  Lett.} {\bfseries 121} (2018) 242001},
  [\href{https://arxiv.org/abs/1808.02077}{{\ttfamily 1808.02077 [hep-ph]}}].

\bibitem{Wang:2017qyg}
W.~Wang, S.~Zhao and R.~Zhu, \emph{{Gluon quasidistribution function at one
  loop}}, \href{http://dx.doi.org/10.1140/epjc/s10052-018-5617-3}{\emph{Eur.
  Phys. J. C} {\bfseries 78} (2018) 147},
  [\href{https://arxiv.org/abs/1708.02458}{{\ttfamily 1708.02458 [hep-ph]}}].

\bibitem{Wang:2019tgg}
W.~Wang, J.-H. Zhang, S.~Zhao and R.~Zhu, \emph{{Complete matching for
  quasidistribution functions in large momentum effective theory}},
  \href{http://dx.doi.org/10.1103/PhysRevD.100.074509}{\emph{Phys. Rev. D}
  {\bfseries 100} (2019) 074509},
  [\href{https://arxiv.org/abs/1904.00978}{{\ttfamily 1904.00978 [hep-ph]}}].

\bibitem{Balitsky:2019krf}
I.~Balitsky, W.~Morris and A.~Radyushkin, \emph{{Gluon pseudo-distributions at
  short distances: Forward case}},
  \href{https://doi.org/10.1016/j.physletb.2020.135621}{\emph{Phys. Lett. B}
  {\bfseries 808} (2020) 135621},
  [\href{https://arxiv.org/abs/1910.13963}{{\ttfamily 1910.13963 [hep-ph]}}].

\bibitem{Dorn:1980hs}
H.~Dorn and E.~Wieczorek, \emph{{Renormalization and short distance properties
  of string type equations in {QCD}}},
  \href{http://dx.doi.org/10.1007/BF01554111}{\emph{Z. Phys. C} {\bfseries 9}
    (1981) 49--57}; erratum:
  \href{https://doi.org/10.1007/BF01410670}{\emph{Z. Phys. C} {\bfseries 9}
    (1981) 274}.

\bibitem{Dorn:1981wa}
H.~Dorn, D.~Robaschik and E.~Wieczorek, \emph{{Renormalization and Short
  Distance Properties of Gauge Invariant Gluoinum and Hadron Operators}},
  \href{http://dx.doi.org/10.1002/andp.19834950208}{\emph{Annalen Phys.}
  {\bfseries 40} (1983) 166--180}.

\bibitem{Braun:2018brg}
V.~M.~Braun, A.~Vladimirov and J.-H.~Zhang, \emph{{Power corrections and
  renormalons in parton quasidistributions}},
  \href{http://dx.doi.org/10.1103/PhysRevD.99.014013}{\emph{Phys. Rev. D}
  {\bfseries 99} (2019) 014013},
  [\href{https://arxiv.org/abs/1810.00048}{{\ttfamily 1810.00048 [hep-ph]}}].

\bibitem{Dotsenko:1979wb}
V.~S.~Dotsenko and S.~N.~Vergeles, \emph{{Renormalizability of phase factors in
  non-abelian gauge theory}},
  \href{http://dx.doi.org/10.1016/0550-3213(80)90103-0}{\emph{Nucl. Phys. B}
  {\bfseries 169} (1980) 527--546}.

\bibitem{Grozin:1994hd}
  A.~G.~Grozin, \emph{{Methods of Calculation of Higher Power Corrections in
      QCD}},
  \href{http://dx.doi.org/10.1142/S0217751X95001674}{\emph{Int. J. Mod. Phys. A}  {\bfseries 10} (1995) 3497--3529},
  [\href{https://arxiv.org/abs/hep-ph/9412238}{{\ttfamily hep-ph/9412238}}].
  
\bibitem{Eidemuller:1997bb}
M.~Eidemuller and M.~Jamin, \emph{{QCD field strength correlator at the
  next-to-leading order}},
  \href{http://dx.doi.org/10.1016/S0370-2693(97)01352-X}{\emph{Phys. Lett. B}
  {\bfseries 416} (1998) 415--420},
  [\href{https://arxiv.org/abs/hep-ph/9709419}{{\ttfamily hep-ph/9709419}}].

\bibitem{Gorishnii:1983su}
S.~G.~Gorishnii, S.~A.~Larin and F.~V.~Tkachov, \emph{{The algorithm for OPE coefficient
  functions in the MS scheme}},
  \href{http://dx.doi.org/10.1016/0370-2693(83)91439-9}{\emph{Phys. Lett. B}
  {\bfseries 124} (1983) 217--220}.

\bibitem{Gorishnii:1986gn}
S.~G.~Gorishnii and S.~A.~Larin, \emph{{Coefficient functions of asymptotic operator
  expansions in minimal subtraction scheme}},
  \href{http://dx.doi.org/10.1016/0550-3213(87)90283-5}{\emph{Nucl. Phys. B}
  {\bfseries 283} (1987) 452--476}.

\bibitem{Vermaseren:2000nd}
J.~A.~M. Vermaseren, \emph{{New features of FORM}},
  \href{https://arxiv.org/abs/math-ph/0010025}{{\ttfamily math-ph/0010025}}.

\bibitem{vanRitbergen:1998pn}
  T.~van~Ritbergen, A.~N.~Schellekens and J.~A.~M.~Vermaseren, \emph{{Group
      Theory Factors for Feynman Diagrams}},
  \href{https://doi.org/10.1142/S0217751X99000038}{\emph{Int. J. Mod. Phys. A}
  {\bfseries 14} (1999) 41--96},
  [\href{https://arxiv.org/abs/hep-ph/9802376}{{\ttfamily hep-ph/9802376}}].

\bibitem{KlubergStern:1974rs}
H.~Kluberg-Stern and J.~B.~Zuber, \emph{{Ward identities and some clues to the
  renormalization of gauge-invariant operators}},
  \href{http://dx.doi.org/10.1103/PhysRevD.12.467}{\emph{Phys. Rev. D}
  {\bfseries 12} (1975) 467--481}.

\bibitem{Collins:1976yq}
J.~C.~Collins, A.~Duncan and S.~D.~Joglekar, \emph{{Trace and dilatation
  anomalies in gauge theories}},
  \href{http://dx.doi.org/10.1103/PhysRevD.16.438}{\emph{Phys. Rev. D} {\bfseries
  16} (1977) 438--449}.

\bibitem{Nielsen:1977sy}
N.~K. Nielsen, \emph{{The energy-momentum tensor in a non-Abelian quark gluon
  theory}}, \href{http://dx.doi.org/10.1016/0550-3213(77)90040-2}{\emph{Nucl.
  Phys. B} {\bfseries 120} (1977) 212--220}.

\bibitem{Spiridonov:1984br}
V.~P.~Spiridonov, \emph{{Anomalous Dimension of $G^2_{\mu\nu}$ and $\beta$
  Function}}, Report No.\ IYaI-P-0378.

\bibitem{Spiridonov:1988md}
V.~P. Spiridonov and K.~G. Chetyrkin, \emph{{Nonleading mass corrections and
  renormalization of the operators $m \bar{\psi}\psi$ and ${G}_{\mu\nu}^2$}},
  {\emph{Yad. Fiz.} {\bfseries 47} (1988) 818--826}
  [{\emph{Sov. J. Nucl. Phys.} {\bfseries 47} (1988) 522--527}].

\bibitem{Chetyrkin:1994ex}
  K.~G.~Chetyrkin and J.~H.~K\"uhn, \emph{{Quartic mass corrections to
      $R_{\mathrm{had}}$}},
  \href{http://dx.doi.org/10.1016/0550-3213(94)90605-X}{\emph{Nucl. Phys. B}
  {\bfseries 432} (1994) 337--350},
  [\href{https://arxiv.org/abs/hep-ph/9406299}{{\ttfamily hep-ph/9406299}}].

\bibitem{Chetyrkin:2018avf}
P.~A. Baikov and K.~G. Chetyrkin, \emph{{QCD vacuum energy in 5 loops}},
  \href{http://dx.doi.org/10.22323/1.290.0025}{\emph{PoS} {\bfseries
  RADCOR2017} (2018) 025}.

\bibitem{Tarasov:1980au}
  O.~V.~Tarasov, A.~A.~Vladimirov and A.~Yu.~Zharkov, \emph{{The Gell-Mann--Low
      function of QCD in the three-loop approximation}},
  \href{http://dx.doi.org/10.1016/0370-2693(80)90358-5}{\emph{Phys. Lett.}
  {\bfseries 93B} (1980) 429--432}.

\bibitem{Larin:1993tp}
  S.~A.~Larin and J.~A.~M.~Vermaseren, \emph{{The three-loop QCD
      $\beta$-function and anomalous dimensions}},
  \href{http://dx.doi.org/10.1016/0370-2693(93)91441-O}{\emph{Phys. Lett. B}
  {\bfseries 303} (1993) 334--336},
  [\href{https://arxiv.org/abs/hep-ph/9302208}{{\ttfamily hep-ph/9302208}}].

\bibitem{Tarrach:1980up}
  R.~Tarrach, \emph{{The pole mass in perturbative QCD}},
  \href{http://dx.doi.org/10.1016/0550-3213(81)90140-1}{\emph{Nucl. Phys. B}
  {\bfseries 183} (1981) 384--396}.

\bibitem{Nachtmann:1981zg}
  O.~Nachtmann and W.~Wetzel, \emph{{The $\beta$-function for effective quark
      masses to two loops in QCD}},
  \href{http://dx.doi.org/10.1016/0550-3213(81)90278-9}{\emph{Nucl.\ Phys.\ B}
  {\bfseries 187} (1981) 333--342}.
  
\end{thebibliography}

\providecommand{\href}[2]{#2}\begingroup\raggedright\endgroup

\end{document}